\documentclass{article}

     \usepackage[preprint,nonatbib]{neurips_2020}

\usepackage[utf8]{inputenc} 
\usepackage[T1]{fontenc}    
\usepackage{hyperref}       
\usepackage{url}            
\usepackage{booktabs}       
\usepackage{amsfonts}       
\usepackage{nicefrac}       
\usepackage{microtype}      
\usepackage[backend=biber]{biblatex}
\usepackage{makecell}
\usepackage{multicol}
\usepackage{caption}
\usepackage{float}
\usepackage{wrapfig}
\usepackage{graphicx}
\graphicspath{ {./figures/} }

\usepackage{tabularx,ragged2e}
\newcolumntype{L}{>{\RaggedRight\arraybackslash\hspace{0pt}}X}
\newcolumntype{P}[1]{>{\RaggedRight}p{#1}}

\title{Encrypted, Anonymized System for Protected Health Information Verification Built via Proof of Stake}

\author{%
  Houjun Liu \\
  The Nueva School\\
  131 E. 28th Ave., \\
  San Mateo, CA 94403 \\
  \texttt{houliu@nuevaschool.org} \\
}

\begin{document}

\maketitle

\begin{abstract}
Digital Health Passes (DHP), systems of digitally validating quarantine and vaccination status such as the New York IBM Excelsior Pass, demonstrate a lawful means to approach some benefits offered by “true elimination” treatment strategies---which focus on the complete elimination of cases instead of investing more in controlling the progression of the disease—of COVID-19. Current implementations of DHPs require region-based control and central storage of Protected Health Information (PHI)---creating a challenge to widespread use across different jurisdictions with incompatible data management systems and a lack of standardized patient privacy controls. In this work, a mechanism for decentralized PHI storage and validation is proposed through a novel two-stage handshaking mechanism update to blockchain proof-of-stake consensus. The proposed mechanism, when used to support a DHP, allows individuals to validate their quarantine and testing universally with any jurisdiction while allowing their right of independent movement and the protection of their PHI. Implementational details on the protocol are given, and the protocol is shown to withstand a $1\%$ disturbance attack at only $923$ participants via a Monte-Carlo simulation---further validating its stability.
\end{abstract}

\section{Introduction}
When correctly implemented \autocite{grout2021}, “elimination” based COVID-19 treatment remains one of the most optimal strategies to rapidly preventing pandemic diseases \autocite{baker2020}. Throughout 2021, countries that continually implemented an elimination-first COVID-19 prevention tactic such as China, Australia, and New Zealand have shown more frequent and rapid returns to “normalcy” compared to those that leveraged a strategy of mitigation \autocite{lu2021}. 

Elimination-based COVID treatment strategies have been shown, if implemented in the United States, to support higher degrees of community well-being \autocite{helliwell2021}. Yet, because non-pharmaceutical COVID-19 elimination interventions (NPIs) typically involves the creation of a continuously operating surveillance scheme with demographic-based targeting and testing, the legal and social challenges to implementing COVID Elimination \autocite{parmet2020} renders it nearly being impossible in terms of policy.

Digital Health Passes (DHP), systems of digitally validating quarantine and vaccination status such as the New York IBM Excelsior Pass, demonstrates a lawful means to approach some benefits offered by true elimination strategies \autocite{gostin2021}. Nevertheless, current implementations of DHPs required region-based centralized control and storage of PHI—creating a challenge to widespread implementation.

In this work, a mechanism for decentralized PHI storage and validation is proposed through the implementation of a novel two-stage handshaking mechanism to the blockchain proof-of-stake consensus mechanism. The proposed mechanism, when used to support a DHP, allows individuals (``agents'') to validate their fitness for societal interaction (e.g. testing, vaccination, etc.) in an elimination-based quarantine while allowing their right of independent movement and the protection of their PHI.

\section{Background}
\subsection{COVID-19 ``Elimination'' Based Strategies}

Based on the work of \textit{Baker, et.al.}\autocite{baker2020}, the overall goal of covid ``elimination'' based strategies is to control a pandemic in a community with a goal of zero in-community transmission or infection.

\begin{figure}[h!]
\begin{tabular}{ccc}
  \begin{minipage}[t]{0.3\textwidth}
    \textbf{Acceptance} \smallskip

    \textit{Goal}  Accepting and profiling novel community members of unknown health status into community using elimination-based protocol.\smallskip

    \textit{NPIs}  Land borders, quarantine facilities, supervised isolation, travel restrictions.\smallskip

    \textit{OxCGRT\textsuperscript{*} Max Value}  5 
  \end{minipage} \hspace{0.5cm}
  \begin{minipage}[t]{0.3\textwidth}
    \textbf{Surveillance} \smallskip

    \textit{Goal}  Surveying the community to appropriately target protected and infected populations, and to “provide reasonable certainty” to case count.\smallskip

    \textit{NPIs}  Multi-modal and high-volume testing, questionnaires, rapid reporting tools.\smallskip

    \textit{OxCGRT\textsuperscript{*} Max Value}  7 
  \end{minipage} \hspace{0.5cm}
  \begin{minipage}[t]{0.3\textwidth}
    \textbf{Removal} \smallskip

    \textit{Goal}  Positive results detected from surveillance are removed from the community to prevent further spread.\smallskip

    \textit{NPIs}  Mandatory and self-isolation, school and facility closures, gathering restrictions.\smallskip

    \textit{OxCGRT\textsuperscript{*} Max Value}  14 
  \end{minipage} 
\end{tabular}

\bigskip
* The Oxford COVID-19 Government Response Tracker (OxCGRT) \autocite{hale2021} is a survey metric of government response tools to the COVID pandemic
\end{figure}
\vspace{-0.75em}
\subsection{(Digital) Health Passes}
In the three-pronged model to “elimination”-based COVID treatment shown above, individually-issued health passes serve as the central registry to keep record of one’s progression through the scheme.

In some jurisdictions (e.g. Australia) each stage issues their own certification—making the “health pass” consist of a collection of documents certifying one’s status. In others, the “health pass” (e.g. New York and California) is fully centrally managed—leading to them having a single point of failure and, as in the case of CDPH Digital Vaccine Record, sometimes containing faulty information.

To make an elimination-based strategy successful, however, a robust health pass system must be in place to clarify and direct the status of an individual in the system. Therefore, building such a Health Pass—adapted to the unique sociopolitical climate in the United States—is critical for an elimination-based policy to be adopted in the US.

\section{A Blockchain Digital Health Pass}

\subsection{Failure of Classical DHPs}
Traditional DHP solutions employ locale-specific validators who take the responsibility of managing the actual PHI supporting their issued DHP. This approach, however, leads to an inaccessibility of information: each jurisdiction has to perform their independent cataloging and, because of the sensitive nature of the PHI typically contained in DHP, have no convenient means by which DHP can be shared between justifications (which are often only the size of a State or even a County in the United States).

A corollary of the above renders current implementations of DHPs in the United States unreliable. If the ``ultimate'' database of trust for a DHP under a jurisdiction has a cataloging mistake (i.e. similar to that aforementioned in California) or—perhaps worse yet—fails to stay online, the DHP looses its function: failing the near-constant stability requirement of true elimination-based COVID prevention strategies.

\subsection{Leveraging Blockchain to Support DHP}
These challenges in stability, distribution, and privacy reflect very similarly to the design goals of blockchain. Blockchain-based solutions have previously been shown to support function such as identity verification \autocite{jamal2019}. State-machine based blockchain implementations such as Ethereum offers a shared, mutually-consensual state between all participants (``nodes'') from which information is stored. 

This state is guaranteed to be the same across all nodes and is viewable by all nodes. Nodes affect change to the global, shared state by executing ``smart contracts'' which is ``validated'' by special nodes guaranteeing the integrity of the shared state through a ``consensus algorithm'' \autocite{buertin2014}.

\subsection{Building a Blockchain DHP}
\begin{wrapfigure}{r}{0.45\textwidth}
  \centering
  \includegraphics[width=0.45\textwidth]{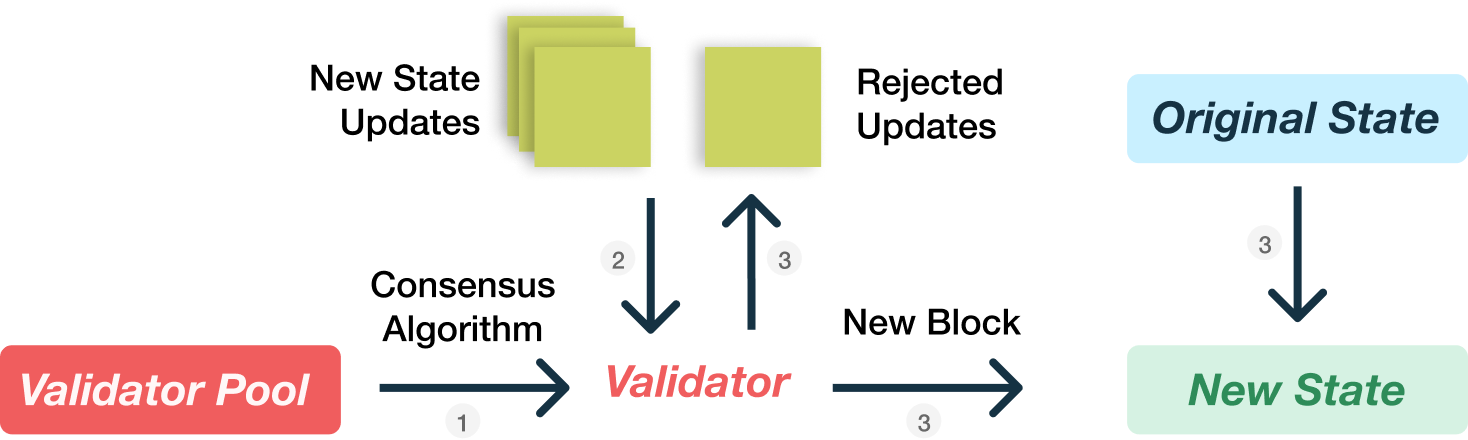}
  \caption{generic state-machine blockchain implementation}
  \label{fig:basicprotocol}
\end{wrapfigure}

State-machine blockchains---as shown in figure \ref{fig:basicprotocol}---can easily represent the claims to support a DHP: each node’s requisite information (e.g. COVID testing status, Vaccination status, etc.) can be stored a part of the shared state, permanent and viewable by all other participants signed by a private key. A user (``agent''), when prompted to represent their identity, can simply leverage the same key to verify their identity.

On the other hand, the process of verifying the purported claims before they are logged in the blockchain to support a DHP—especially given the amount of PHI involved in the process—is a much more complex one. Instead of simply evaluating the mathematical validity of an algorithmic contract as in the case for Ethereum, DHP Blockchain validators have to review documents supporting a state-update claim and ensure their authenticity.

The proposed DHP Blockchain protocol therefore forsakes the traditional Proof-of-Work and Stake algorithms for consensus and presents a new multi-stage human-in-the-loop validation and consensus mechanism that provides a means to verify DHP claims’ supporting documentation while still guaranteeing practical anonymity at scale.

\subsection{Considerations of Deploying Blockchain DHP}
Blockchain DHPs provide a decentralized and stable solution to problems presented by Classic DHPs by foregoing the need for a central controlling server/party. Although Blockchain DHPs solve many of such issues, they also provide a differing set of challenges. Verification of Blockchain DHPs are done by multiple untrusted partners (as opposed to the one central point-of-trust by Classic DHPs) which means that they are—like all other Blockchain applications—at risk of disturbance. 

The necessity for multiple machines to network and validate identity/health claims in conjunction renders the proposed solution susceptible to multiple, collaborating disturbance agents which in conjunction overwhelm the factual verification capabilities of the network. 
The risk of such disturbance attacks, along with the necessary agent population needed to prevent the verification claims made by disturbance agents from being successful, is quantified in the section below with a probability model.

\section{Implementation Details}
\begin{figure}
  \centering
  \includegraphics[width=0.85\textwidth]{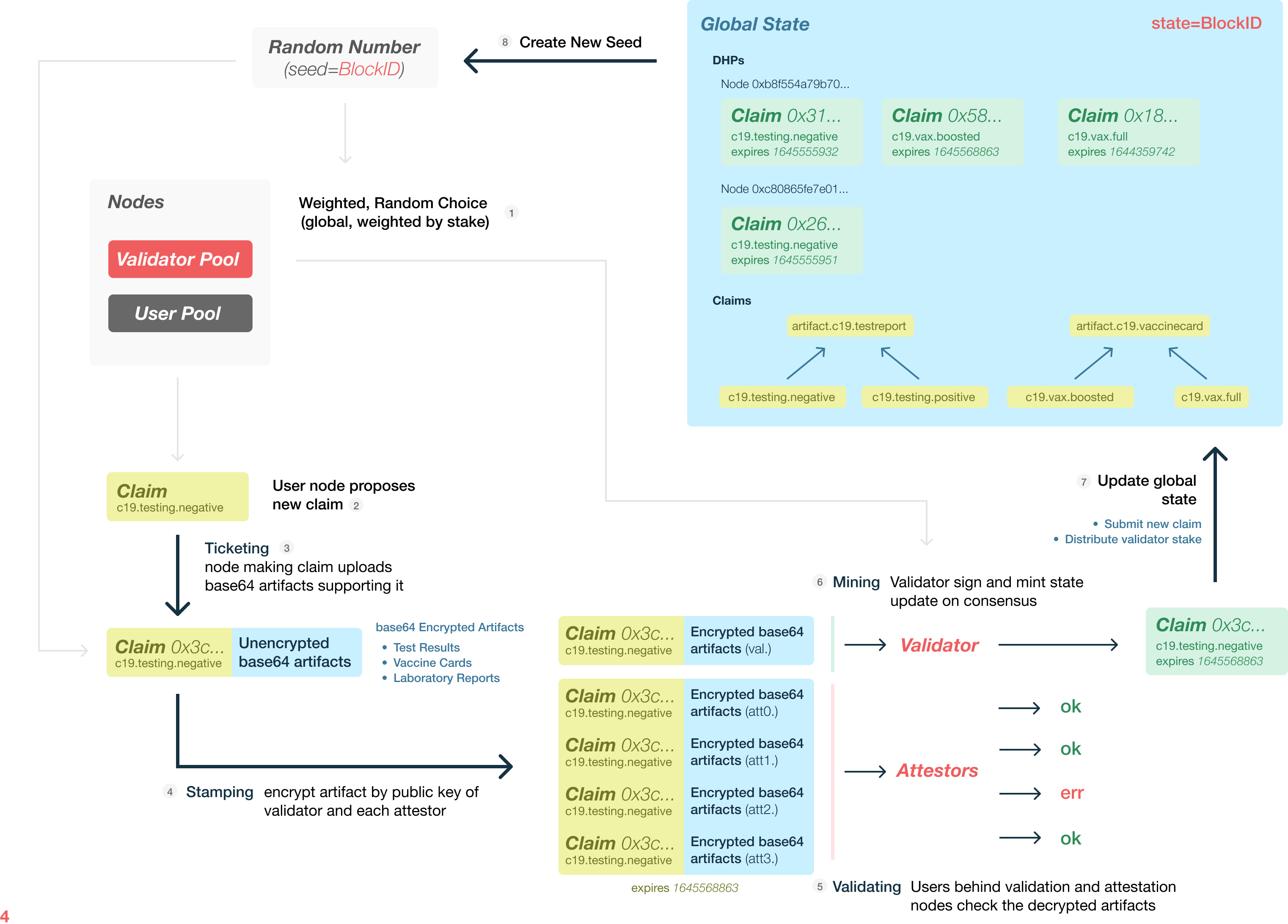}
  \caption{proposed protocol}
  \label{fig:protocol}
\end{figure}

Figure \ref{fig:protocol} on \pageref{fig:protocol} illustrates the entirely of the proposed verification cycle. Futher implementation details are here examined.

\subsection{Validators, Attestors, and Stake}
The system that upholds the integrity of claim making in the proposed algorithm works in the same manner as that in traditional proof-of-stake. A fiat currency is circulated throughout nodes wishing to participate in validating; to be eligible for validation, a validator/attestor must ``stake'' a certain unit of the currency.

The ``validator'', scaled by the proportion of their stake, is selected randomly based on a pseudo-random global seed by the current state. Four other ``attestors'', scaled also by their stake, are selected to form a “committee” to validate each block of transactions. 

\subsection{Claims and Claim Ticketing}
Each non-validator node, primarily, operate by proposing new ``claims''. Claims are unverified statements (such as updated COVID testing status) lodged by nodes and are represented by a string ID corresponding to a claim whose string criteria are previously placed in the publicly viewable global state. 

To make a claim, the proposing node has to first ``ticket'' the claim. This is a process by which the node takes a claim and appends to it base64 encoded artifacts (such as lab reports) to support it. The whole package (claim + artifacts) is then hashed in order to check for tampering later.

\subsection{Ticket Stamping}
Ticketed claims are then ``stamped'': the base64 supporting information to each claim is encrypted using the public keys to the validator and attestors. As the blockchain mechanism itself grantees global state concordance, the proposing node can simply use the same randomness seed in the global state—making its choices definitively agree with that of any other participant of the chain. The encrypted results, along with the proposed claim, is published to all peers.

\subsection{Ticket Validation}
The chosen validators and attestors will, upon receipt, decode the claim and decrypt the supporting information. Human evaluators will here make a judgment of whether the documents support the claim being made.

The chosen ``validator'' node, upon successful human evaluation, will mint a new signed ``block'' containing the claims made and the corresponding nodes that made the claims. Before the block is accepted, however, the attestors will ensure that the majority of their judgments of validity is equivalent to that of the validator. If this is indeed true, the block is accepted. If not, the validator would—as with traditional proof-of-stake—loose their stake value and right to validate.

\subsection{Reference Implementation}
All encryption was performed through AES-256 PGP; inter-peer communication was implemented through tcp using ZeroMQ. The codebase is written using Clojure (v1.10.1), running the Java SE 14 (HotSpot) Virtual Machine. Risk modeling was performed using SageMath 9.4 upon Python 3.10.1.

\section{Risk of Disturbance}
The 4-attestor model of the proposed algorithm exhibits the same weakness towards attacks than that of the traditional Proof-of-Stake model employed by the Ethereum blockchain verification \autocite{buertin2014}. Hence, the model proposed must be verified with the same means of quantifying risk to identity a minimally-viable population.

Take some $N$, population sample, $R_v$, proportion of population to validator/attestor pool, and $R_d$, rate of “disturbance” upon the validator/attestor pool, a compound probability $P_{nd}$ can be derived representing the probability of the block randomly sampling dependently 0, 1, or 2 of the disturbance agents (and therefore rendering the faulty agents’ attestation of no effect.) The chance of disturbance, $P_d$, is therefore the inverse of $P_{nd}$.

\begin{figure}[h!]
\begin{tabular}{cc}
  \begin{minipage}[b]{0.62\textwidth}
    \includegraphics[width=0.95\textwidth]{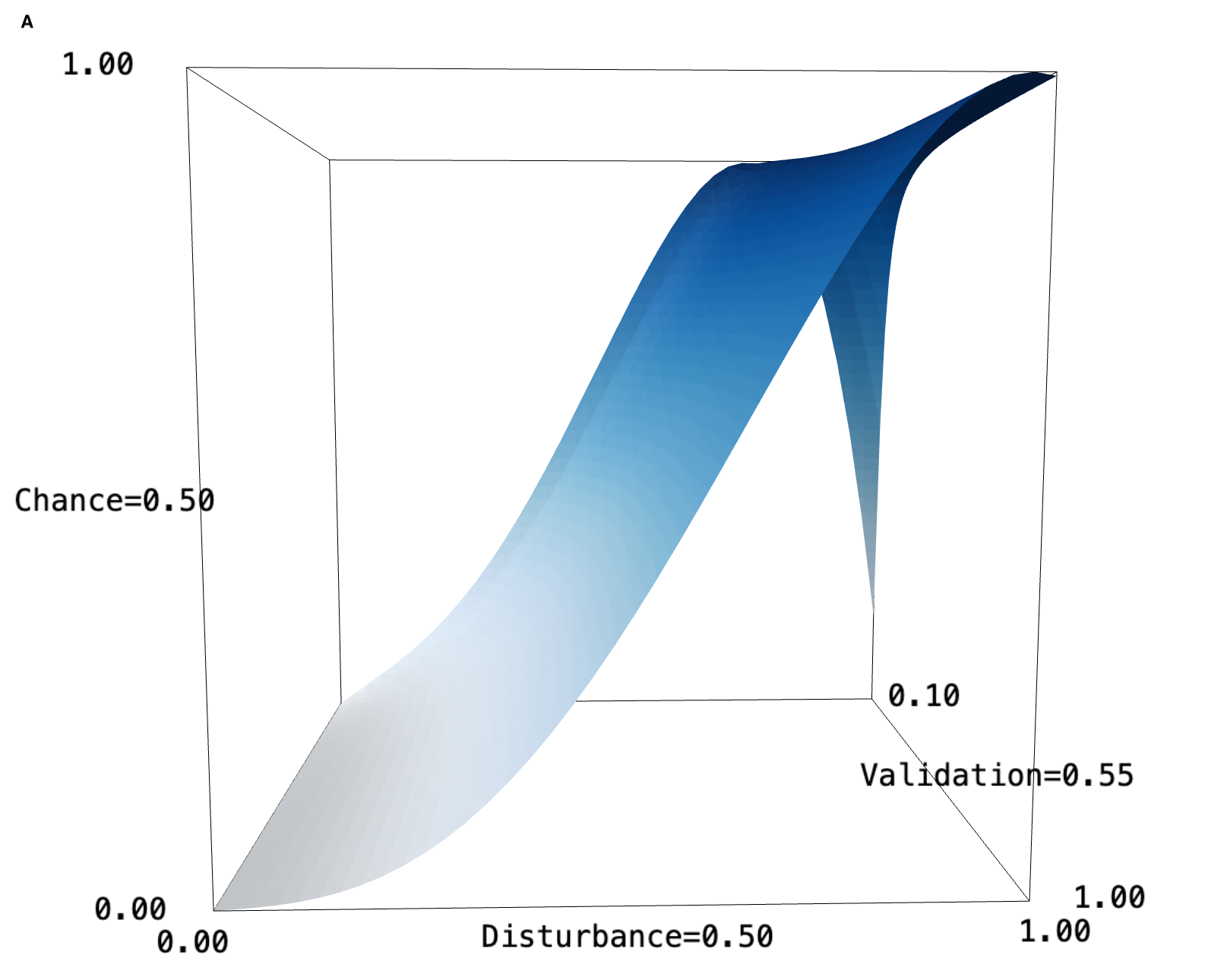}

    \medskip
    \small
    \textbf{A} cumulative disturbance probability at $N=100$ of chance of disturbance as a variable of validation and disturbance populations. \textbf{B} chance of disturbance given $R_y=R_d=0.1$ at various $N$. \textbf{C} $R_y$ required for $R_d<0.01$ at various $N$. 
  \end{minipage} \hspace{0.1cm}
  \begin{minipage}[b]{0.38\textwidth}
    \includegraphics[width=0.95\textwidth]{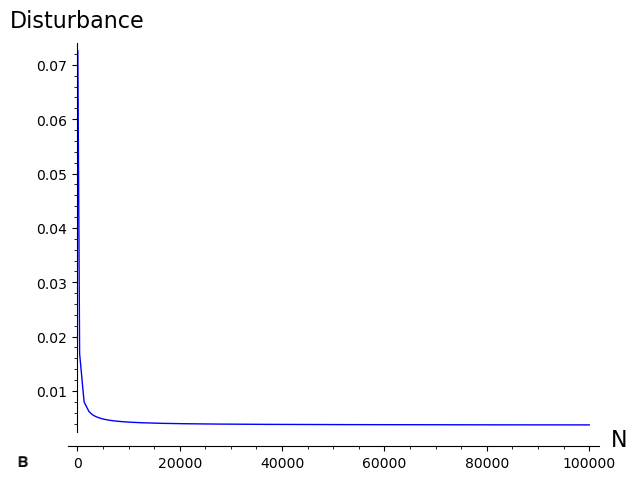}\\
    \includegraphics[width=0.95\textwidth]{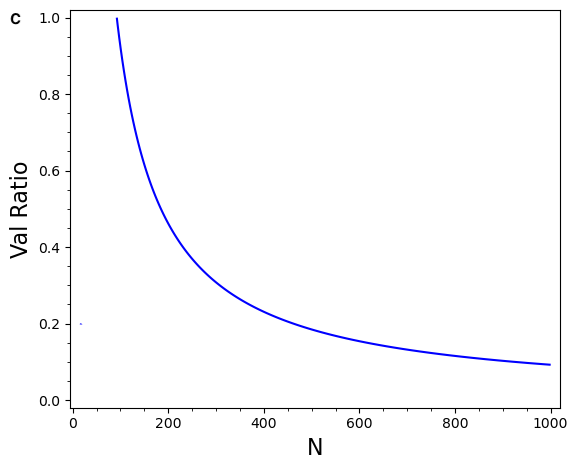}\\
  \end{minipage} 
\end{tabular}
\caption{chance of disturbance as modeled by various parameters.}
\label{fig:disturbance}
\end{figure}

Relative risk is controlled $<0.01$ at large $N > 923$ at $R_y=R_d=0.1$. According to the risk model aforementioned, the $Pd \propto N$. Therefore, at $N>923$, $Pd<0.01$. At smaller values of $N$, the risk model indicates the need for a higher validator election ratio from the agent pool to control the chance of disturbance at $<0.01$. Figure \ref{fig:disturbance} highlights the relative risk as well as the necessary chance of validator election at various $N$.

\section{Conclusion and Discussion}
In this work, a novel mechanism introducing asymmetrically-encrypted verification artifacts built atop the existing proof-of-stake scheme for blockchain is proposed to enable blockchain-based Protected Health Information (PHI) verification. 

In the proposed scheme, 4 attestors and 1 validator are selected based on stake which performs the block minting and result attestation of PHI claims. PHI is made available only to the selected validation panel, and the encrypted artifacts are never stored on the chain. 
Similar to the proof-of-stake “validation panel” design, this scheme reduces the possibility of noise and is self-correcting of outlying errors. The duplicity and decentralization of Blockchain also allows for widespread distribution of the verified health information in a serverless manner that does not require a central point of trust and stability.

As is standard on Blockchain applications, verified user information (``claims'') are also kept in identity-less hashed key fingerprints much like ``wallets'' in traditional cryptocurrency applications. Hence, the private signing key of the user submitting claims acts naturally as a Digital Health Pass (DHP)—selectively identifying and de-anonymizing only the specific claims made by the user holding the private key when they wish.

Like all other trustless, consensus-paged algorithms, the scheme proposed—when deployed on a live blokchain—is susceptible to overwhelming disturbance. As such, a probabilistic model is created to quantify the chance of the scheme creating a false assurance. According to the modeled result, the chance of the scheme being disturbed is $<1\%$ when user count exceeds just 923 active participants under a $10\%$ validation pool. At smaller user counts of around 200, the disturbance chance can still be controlled at $<1\%$ if the validation pool is increased to a large minority of the user population at around $40\%$.

This model indicates that the proposed algorithm can sufficiently serve as a stable and distributed alternative to centralized PHI verification: creating a safe alternative that addressed the accessibility, privacy, and stability issues with conventional DHPs currently deployed. Therefore, with the reference implementation serving as the core of the service, and the necessary regulatory approval, it is possible to bring this service to use as a DHP in localities in the short-term future. 

As the COVID pandemic eases, this service can also aid in the rapid, stable, and large-scale anonymous validation of health conditions such as allergies, diabetes, or other PHI-based facts: creating numerous applications in infrastructure, enterprise, education, and industry.

\begin{ack}
  The author acknowledges Mr. Wesley Chao of The Nueva School and Dr. Xin Liu at the University of California, Davis for their technical and implementation guidance in the protocol. The author would further like to acknowledge Dr. John Feland of CableLabs, inc., Dr. Mark Hurwitz of The Nueva School, and Mr. Sean Cheong at the University of California, Berkeley for their feedback on the protocol and disturbance calculations.
\end{ack}

\printbibliography

\end{document}